# Interaction-induced insulating state in thick multilayer graphene

Youngwoo Nam[1], Dong-Keun Ki[1], Mikito Koshino[2], Edward McCann[3], and Alberto F. Morpurgo[1]

[1] Department of Quantum Matter Physics (DQMP) and Group of Applied Physics (GAP), University of Geneva, 24 Quai Ernest-Ansermet, CH1211 Genéve 4, Switzerland

[2] Department of Physics, Tohoku University, Sendai 980–8578, Japan

[3] Department of Physics, Lancaster University, Lancaster, LA1 4YB, UK

To whom correspondence should be addressed. Email: Dongkeun.Ki@unige.ch or Alberto.Morpurgo@unige.ch

**Abstract**

Close to charge neutrality, the low-energy properties of high-quality suspended devices based on atomically thin graphene layers are determined by electron-electron interactions. Bernal-stacked layers, in particular, have shown a remarkable even-odd effect with mono- and tri-layers remaining gapless conductors, and bi- and tetra-layers becoming gapped insulators. These observations –at odds with the established notion that (Bernal) trilayers and thicker multilayers are semi-metals– have resulted in the proposal of a physical scenario leading to a surprising prediction, namely that even-layered graphene multilayers remain insulating irrespective of their thickness. Here, we present data from two devices that conform ideally to this hypothesis, exhibiting the behavior expected for Bernal-stacked hexa and octalayer graphene. Despite their large thickness, these multilayers are insulating for carrier density $|n|<2\text{-}3\times 10^{10}$ cm$^{-2}$, possess an energy gap of approximately 1.5 meV at charge neutrality –in virtually perfect agreement with what is observed in bi and tetra layer graphene– and exhibit the expected integer quantum Hall effect. These findings indicate the soundness of our basic insights on the effect of electron interactions in Bernal graphene multilayers, show that graphene multilayers exhibit unusual and interesting physics that remains to be understood, and pose ever more pressing questions as to the microscopic mechanisms behind the semimetallic behavior of bulk graphite.

**Introduction**

The recent study of high-quality suspended Bernal-stacked tetralayer graphene (4LG) has revealed a drastic effect of electronic interactions at zero magnetic field (*B*=0), which turn the system into an insulator in a narrow range of charge density $|n|$<2-3×10$^{10}$ cm$^{-2}$ close to the charge neutrality point (CNP, where *n*=0) [1]. This behavior, similar to the one of suspended bilayers [1-5], differs from that of equally high quality monolayers (1LG) and Bernal-stacked graphene trilayers (3LG) that remain conducting at low temperatures [6-9]. The resulting "even-odd" effect is illustrated in figure 1(a) by the comparison of the temperature (*T*) dependence of the conductivity at the CNP ($\sigma_{min}$) measured in the different *N*-layers. An additional surprise comes from the observation that in 4LG the conductivity at charge neutrality is more strongly suppressed than in 2LG (see figure 1(a)), i.e., the insulating state in 4LG is more pronounced than in 2LG. This is an unexpected, counterintuitive finding, as one would anticipate that –upon increasing thickness– the behavior of multilayer graphene should approach that of bulk graphite, which is a semimetal and remains highly conducting at low *T* [10]. Here we present experimental results that –despite being possibly even more unexpected– validate the physical scenario responsible for the occurrence of the even-odd effect, and show that graphene-based systems are continuing to reveal interesting surprises.

In the absence of a comprehensive microscopic theory, our current understanding of the effect of electron-electron interactions in thick multilayers relies on a phenomenological approach based on the so-called minimal tight-binding model of multilayer graphene, augmented with a staggered layer potential (as described in detail in the supporting information) [1]. The minimal tight-binding description includes only nearest-neighbor intra- and inter-layer coupling (usually denoted by parameters $\gamma_0$ and $\gamma_1$, see figure 1(b)). For even *N*, it results in *N/2* sets of gapless bilayer-like bands with quadratic dispersion whereas, for odd *N*, there are (*N-1)/2* sets of gapless bilayer-like bands plus one pair of linearly-dispersing monolayer-like bands [1, 11-17]. The staggered layer potential $V_i = (-1)^{i+1}\Delta$ (where $i = 1,2,3,...$ is a layer index and $\Delta$ is the order parameter), assumed to originate from a mean-field treatment of electron-electron interactions, is meant to generalize the theoretical description developed for bilayer graphene (e.g., $\Delta$ originates form exchange interaction and its sign depends on spin and valley) [3, 18-23]. For even *N* multilayers, it breaks inversion symmetry, opening a gap at zero density and creating an insulating state (see figure 1(c)). For odd *N* multilayers, it gaps the bilayer-like bands, but not the monolayer-like bands, which remain gapless, so that odd *N* multilayers have a conducting state (this is because the wave functions of the monolayer band only have finite amplitude on every other layer and, thus, they experience the staggered potential as a constant layer potential; see supporting information). This same model also predicts that the most robust integer quantum Hall state in a *N* multilayer should be observed at a filling factor $\nu = nh/eB = 2N$ (*h*: the Planck's constant; *e*: electric charge; figure 1(d)) [15], which is particularly relevant for experiments, as the analysis of the quantum Hall effect sequence provides a way to determine the thickness of the layer investigated [1]. These clear predictions are seemingly in contrast with the known behavior of bulk graphite (which is a conductor at low temperature, and not an insulator) [10], and with the observed properties of devices in which thick Bernal multilayers are in direct contact with a substrate [24-29]. It is therefore crucial to establish whether the scenario proposed to describe the effect of electron interactions remains valid for layers thicker than the thickest ones investigated so far, or whether the insulating behavior of 4LG –which provided the experimental evidence supporting such a scenario– is somehow coincidental.

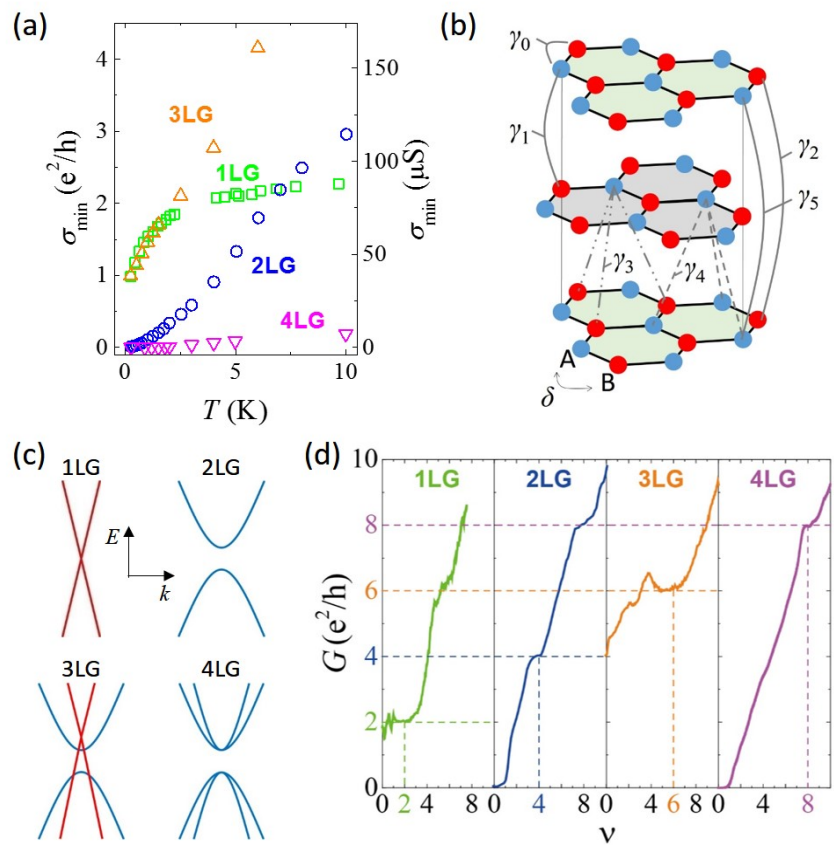

**Figure 1. "Even-odd" interaction effect from monolayer graphene to 4LG. (a)** Temperature dependence of the minimum conductivity ($\sigma_{min}$) at charge neutrality for $N$=1-4 graphene multilayers, illustrating the "even-odd" effect. A strong suppression of $\sigma_{min}$ at low temperatures is observed for even $N$ multilayers (2LG and 4LG), but not for odd-$N$ multilayers (1LG and 3LG), in which even at the lowest temperature $T$=250 mK, $\sigma_{min}$ is a few $e^2/h$ (Data for 1LG, 2LG [30, 32], 3LG, and 4LG [1] are extracted from suspended devices realized in our laboratory.). **(b)** Structure of Bernal-stacked graphene multilayers; the tight-binding hopping parameters of the Slonczewski-Weiss-McClure model [34, 35] are indicated (see Discussion Section). **(c)** Schematic illustration of the band structure of Bernal-stacked 1LG, 2LG, 3LG, and 4LG as predicted by the minimal tight-binding model in the presence of a staggered layer potential, leading to insulating behavior at charge neutrality in even-$N$ multilayers. **(d)** Low magnetic field quantum-Hall effect measured in suspended graphene multilayers with $N$=1-4 realized in our laboratory, exhibiting the first plateau at $v$ =2x$N$, as expected from the theoretical scenario outlined in the main text.

Here we show data measured on two different suspended thick graphene devices that precisely match what is expected for $N$=6 and $N$=8 Bernal stacked multilayers in the presence of interactions, in line with the even-odd scenario. In particular, these devices are insulating for $|n| < 2\text{-}3\times10^{10}$ cm$^{-2}$, have a gap of approximately 1.5 meV comparable to that found in 2LG and 4LG [1-5], and exhibit a clear first integer quantum Hall plateau at filling factors $v$ = 12 and 16 respectively (with the conductance having the correct corresponding quantized value of 12 and 16 $e^2/h$). These findings provide a strong validation that the basic aspects of our understanding of electron interactions in Bernal multilayers are correct (at least up to $N$=8), bring to light some very unusual properties of multilayer graphene, and reiterate more pressingly questions about the microscopic mechanisms responsible for the semimetallic state of bulk graphite. Specifically, if it is the case that at some larger $N$ even Bernal multilayers cease to be insulating, at what thickness does this happen and which physical processes determine that length scale?

**Methods**

Suspended graphene devices were fabricated using polydimethylglutarimide (PMGI)-based lift-off resist (LOR 10A, MicroChem) as a sacrificial layer [30, 31]. Graphene flakes were mechanically exfoliated from natural graphite onto 1 μm-thick LOR covered heavily doped Si substrate (acting as a back-gate) capped with 285 nm-thick SiO$_2$. Thick graphene multilayers were selected based on their strong optical contrasts, and electrical contacts (10 nm Ti/70 nm Au) were made by using a conventional electron-beam lithography and lift-off technique. For a multi-terminal geometry, oxygen plasma etching was used to define the shape of the flake [30]. Lastly, multilayer graphene was suspended by removing LOR resist underneath. Current annealing was performed at 4.2 K in vacuum for several times until the device shows a very sharp resistance peak with a corresponding $\delta n$ ~ 2-3×10$^9$ cm$^{-2}$ which is essential to observe the phenomena discussed here (current annealing is the most critical experimental step in the device fabrication procedure). Once a well-defined resistance peak appears, transport measurements were carried out in various temperature ($T$), magnetic field ($B$), and bias voltage ($V_{sd}$) using a standard low frequency lock-in technique. Two types of device were studied, device A in a multi-terminal geometry which allows us to measure the resistance in different configurations and device B in a more conventional two-terminal geometry.

**Results**

Our experiments rely on suspended graphene devices with extremely high-quality, in which charge inhomogeneity is reduced to well below 10$^{10}$ cm$^{-2}$ as needed to observe the effects of electron interactions: the realization of these devices is technically challenging and so far it has been reported only for 4LG or thinner multilayers [1-9, 30, 32]. By following the procedure that we have employed previously [30, 32], we have succeeded in realizing two high quality suspended devices based on multilayers that –as indicated by their contrast under an optical microscope– are thicker than 4LG. We will hereafter refer to these devices as to A and B. For device A, a four-terminal configuration could be implemented, whereas device B was realized in a two-terminal configuration. The basic aspects of their electrical characterization are illustrated in figure 2 (figures

2(a)-(c) show data for device A and figures 2(d)-(f) show the corresponding measurements for device B).

Figures 2(a) and 2(d) show the dependence of the square resistance of devices A and B measured at $T$ = 4.2 K as a function of gate voltage ($V_g$). An extremely sharp peak around the charge neutrality point is observed. By plotting $\sigma(n)$ in a double-logarithmic scale [6, 7], as shown in the insets of the figures, we find that the peak width corresponds in both cases to very small density inhomogeneity, $\delta n$ ~ 2-3×10$^9$ cm$^{-2}$ (the conversion factor between $V_g$ and $n$ is obtained from the analysis of the quantum Hall effect, as explained later). This value is comparable to what has been reported in the very best suspended graphene devices irrespective of their thickness [1-9, 30, 32], and provides a first indication of the device quality. For both devices, the square resistance at the charge neutrality point is approximately 350 k$\Omega$, comparable to the value measured in 4LG at the same temperature and 10 times larger than the value in 2LG (see figure 1(a)). As we discuss in detail below, this large value of the square resistance is due to the insulating nature of the devices very close to the charge neutrality point.

Devices A and B clearly exhibit the integer quantum Hall effect, whose analysis allows us to characterize their quality, and –as we mentioned above– is important to validate the theoretical scenario for the effect of electron-electron interactions in graphene multilayers [1]. The details of the measurements are different for the two devices, since only one of them is a multi-terminal structure, but in both cases the data enable us to extract the necessary information. Specifically, figure 2(b) shows the longitudinal four-terminal magnetoresistance of device A. It shows clear minima at values of magnetic field $B$ that disperse linearly with $V_g$ fanning out of the origin, a manifestation of Landau level formation. The minima become visible as the magnetic field is increased past $B$ ~ 0.1-0.2 T implying, from the criterion $\mu B$ >1 for their visibility, that the carrier mobility $\mu$ ~ 100 000 cm$^2$/Vs [6, 7]. Similar considerations –and a comparable estimate $\mu$ ~ 100.000 cm$^2$/Vs– can be made for device B on the basis of the data shown in figure 2(e). In this case, since measurements can only be made in a two terminal configuration, the fan diagram is obtained by plotting the derivative of the measured conductance $G$ relative to gate voltage, $dG/dV_g(V_g,B)$.

Figures 2(c) and 2(f) show the magnetotransport data plotted as a function of filling factor $\nu$ =$nh/eB$ for the two devices. The density $n$ is determined from the relation $n=\alpha(V_g-V_{CNP})$. $V_{CNP}$ corresponds to the value of $V_g$ for which the resistance is maximum, and $\alpha$ is obtained by optimizing the collapse of the data for each measured quantity on a single curve (as expected for transport in the quantum Hall regime). Quantum Hall plateaus at integer multiple values of $e^2/h$ are clearly visible both in the transverse conductivity $\sigma_{xy}$ of device A and in the two terminal conductance of device B. For device A, the plateaus occur in concomitance with minima in the longitudinal conductivity $\sigma_{xx}$ (again as it should be for quantum Hall transport). Importantly, the value of the quantized conductivity (conductance) at the first plateau observed for devices A and B is 12$e^2/h$ and 16$e^2/h$, and occurs respectively at filling factors $\nu$ =12 and 16, showing the internal consistency of the experimental results. These values are larger than what is seen in suspended 4LG devices (see figure 1(d)), in which the first, most robust integer quantum Hall state appearing at low field corresponds to $\nu$ = 8 [1]. Finally, the analysis of the data also consistently gives the slope of the dispersing features in the fan diagrams shown in figures 2(b) and 2(d) pointed to by the arrows.

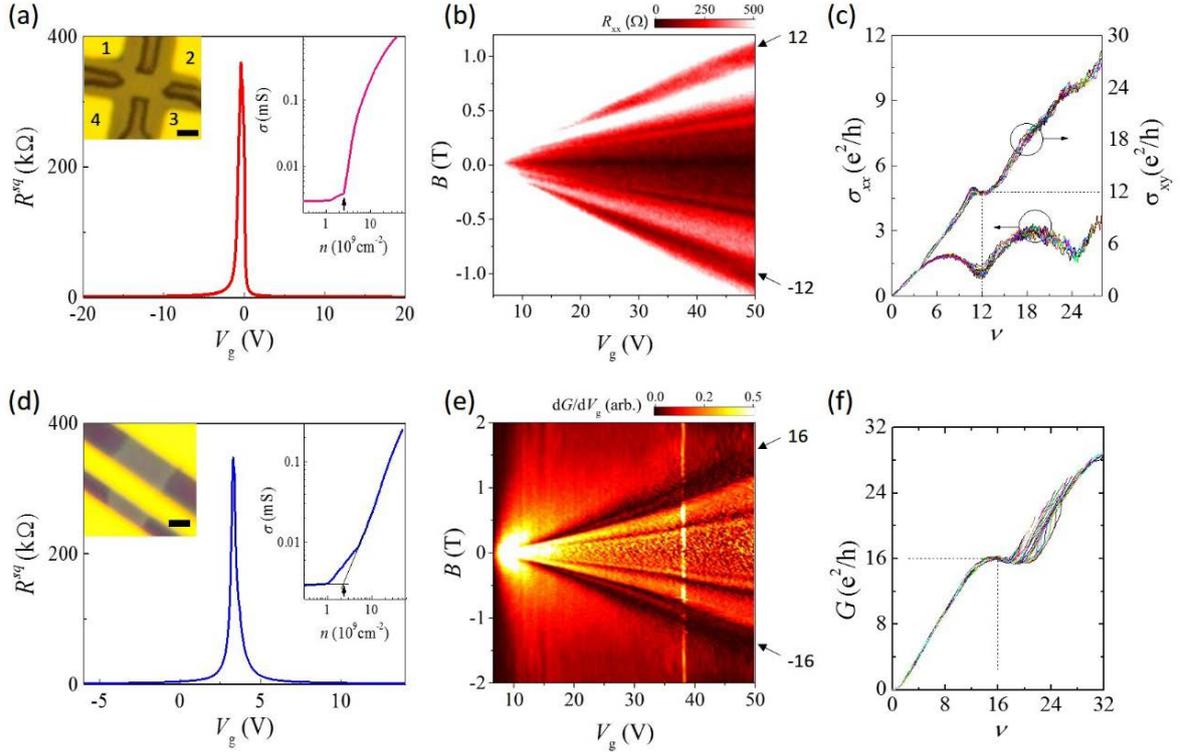

**Figure 2. Gate-dependent low-temperature magneto-transport.** Panels **(a-c)** and **(d-f)** show data from device A and B, respectively. **(a, d)** $V_g$-dependence of the square resistance ($R^{sq}$) measured at 4.2 K in a two-terminal configuration, showing in both cases a peak value of 350 kΩ at the CNP. The right insets show a double-logarithmic plot of the density dependence of the conductivity, $\sigma(n)$, enabling the determination the charge inhomogeneity $\delta n \sim 2 \times 10^9$ cm$^{-2}$ (see the arrows). The left insets show optical microscope images of the devices (the bar is 1 μm). **(b, c)** Quantum-Hall effect in device A (data taken at 250 mK, in a four-terminal configuration). Panel **(b)** shows a color plot of the longitudinal resistance $R_{xx}$ versus $V_g$ and $B$, with the minimum in $R_{xx}$ at $\nu = 12$ (pointed to by the arrow) appearing first at a magnetic field $B$ as low as 0.2 T. Panel **(c)** shows the longitudinal and transverse conductivity plotted as a function of filling factor $\nu$, collapsing together as expected in the quantum Hall regime (data taken for $B$ between 0.3 T and 1.2 T). The occurrence of the first integer quantum Hall state at $\nu = 12$ with $\sigma_{xy} = 12e^2/h$ is apparent. **(e, f)** Quantum Hall effect in device B (data taken at $T = 250$ mK). **(e)** Color-plot of d$G$/d$V_g(V_g, B)$, with a dispersing minimum (pointed to by the arrow) corresponding to a $\nu = 16$ quantum Hall state appearing first at $B \sim 0.1$ T. Panel **(f)** illustrates the expected scaling of the corresponding plateaus at $G = 16e^2/h$ when plotted against $\nu$ (the traces are taken for $B$ between 0.6 T and 1.5 T). For device B, the occurrence of the first integer quantum Hall state at $\nu = 16$ with $\sigma_{xy} = 16\ e^2/h$ is apparent. The $\nu = 12$ and 16 at which the first quantum-Hall plateau appears in device A and B are precisely the values expected for Bernal-stacked 6LG and 8LG.

Having completed the basic device characterization, we discuss the origin of the highly resistive states observed at the charge neutrality point (figures 2(a) and 2(d)). To this end we measure the $V_g$-dependent conductivity for different values of temperature between 250 mK and 40 K, at small charge densities $n$ near charge neutrality. For both device A (figure 3(a)) and B (figure 3(b)) the data show that the conductivity decreases rapidly upon lowering $T$. At the lowest temperature reached in our experiments (250 mK), the conductivity $\sigma_{min}$ at the CNP decreases to only about 0.01 μS (≈3×10$^{-4}$$e^2/h$, limited by our measurement setup). For device A, the multi-terminal configuration also allows us to verify the spatial homogeneity of the highly resistive state [1], by checking that the two-terminal conductance measured between different pairs of contacts exhibits an identical behavior (see the inset of figure 3(a)). The Arrhenius plot, $\sigma_{min}$ (in log-scale) as a function of $1/T$, shown in figure 3(c) demonstrates that transport in this regime is thermally activated in both device A and B. From the fitting $\sigma_{min} \sim \exp(-E_A/2k_BT)$, we extract the activation energy $E_A \approx 15$ K (≈ 1.3 meV), which coincides with the value measured previously in 2LG and 4LG [1, 3-5]. In both device A and B, therefore, the highly resistive state originates from an energy gap present near charge neutrality, whose magnitude and carrier density dependence are remarkably similar to what has been found in suspended bilayer and 4LG (see figure 3(c)).

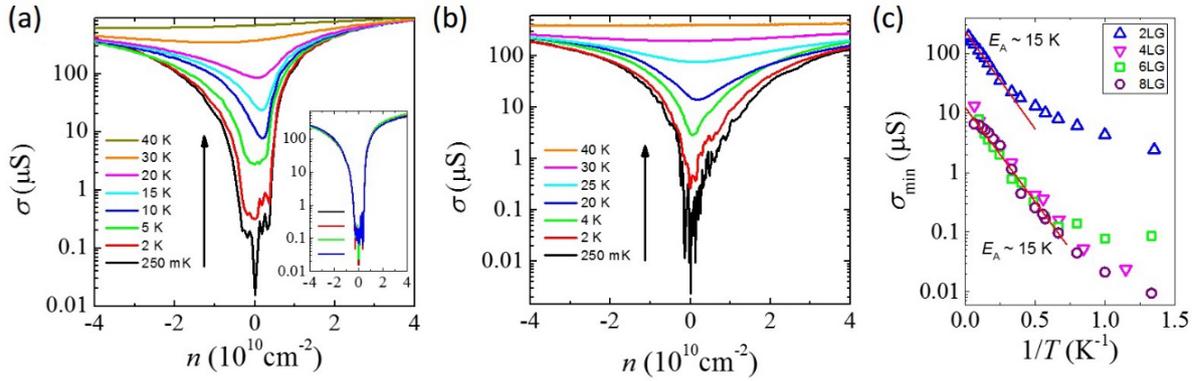

**Figure 3. Insulating state at charge neutrality. (a, b)** Conductivity ($\sigma$) as a function of $n$ measured in device A and B, respectively, exhibiting a strong suppression with decreasing $T$ in a very narrow range of $|n|<2\times10^{10}$ cm$^{-2}$ around the CNP (the inset of (**a**) shows that two-terminal measurements done on device A at $T$ =250 mK using different pairs of contacts give the same result: this observation demonstrates the device homogeneity). Upon increasing $T$, the minimum conductance ($\sigma_{min}$) at charge neutrality rapidly increases in both devices, exhibiting a thermally activated behavior, $\sigma_{min} \sim \exp(-E_A/2k_BT)$. This is shown in panel **(c)**, in which $\sigma_{min}$ is plotted as a function of $1/T$ for devices A (squares) and B (circles), as well as for 2LG (up-triangles) and 4LG (down-triangles). Note that the behavior of device A and B is virtually identical —at a quantitative level– to that of 4LG. Additionally, for all devices, the activation energy seen for $T$ > 1-2 K coincides, i.e., $E_A \sim 15$ K in all cases. Taken together with the observation of the first quantum-Hall plateau at $\nu$ = 12 and 16, therefore, the behavior of device A and B matches precisely the expected behavior for 6LG and 8LG, according to the phenomenological scenario discussed in the main text.

Finally, we look at the behavior of transport at finite bias. In the highly resistive state, the application of a voltage between source and drain electrodes ($V_{sd}$) provides additional evidence for the presence of an energy gap [1, 3, 5, 33]. Figures 4(a) and 4(c) show a color-plot of the differential conductance d$I$/d$V$ of the A and B devices as a function of $V_{sd}$ and $V_g$ measured at 250 mK. In both cases, the data near charge neutrality shows a strongly suppressed conductance (< 0.01 $e^2/h$) in a small bias range, $|V_{sd}|$<1-2 mV. At $|V_{sd}|$ larger than 2 mV – i.e, on an energy scale comparable to the activation energy found from the analysis of the temperature dependence– d$I$/d$V$ is found to increase pronouncedly (see the line cuts at specific $V_g$ in figures 4(b) and 4(d)). We emphasize once again the similarity of this behavior to what has been previously observed in 2LG and 4LG [1, 3, 5].

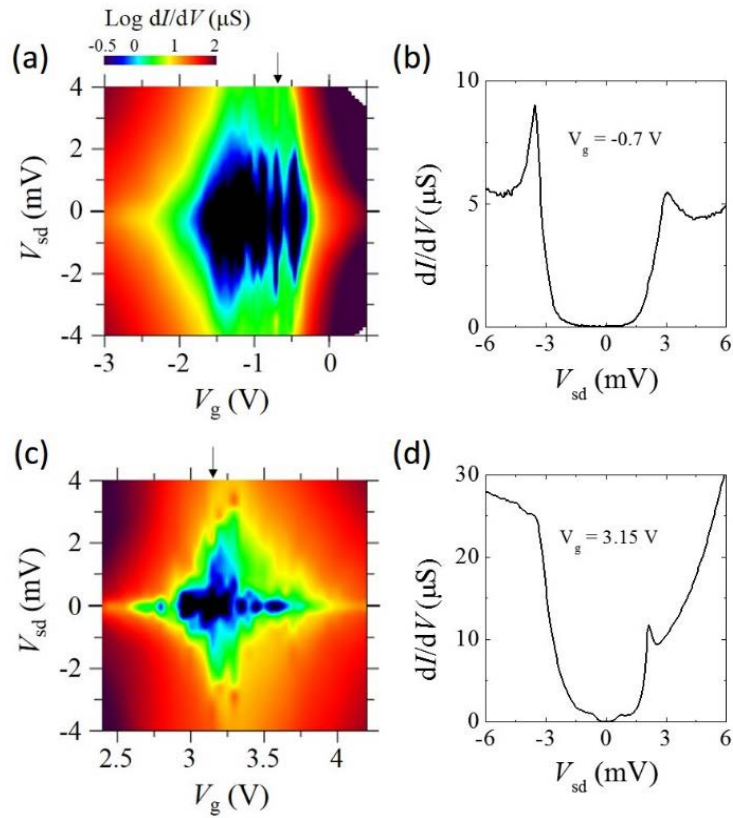

**Figure 4. Bias-dependent transport around charge neutrality.** Panels **(a, b)** and **(c, d)** represent the bias voltage ($V_{sd}$) dependence of the differential conductance d$I$/d$V$ measured at 250 mK on device A and B, respectively, close to the CNP. The color-plots of log(d$I$/d$V$) against $V_g$ and $V_{sd}$ (**a, c**) clearly exhibits a vanishing d$I$/d$V$, within the accuracy of the measurement, for $V_{sd}$ below 1-2 meV in both devices (see the dark blue region). The strong suppression of d$I$/d$V$ is clearly visible in panels **(b, d)** which plot line cuts of the color-plots at $V_g$ = -0.7 V and 3.15 V respectively (as indicated by the arrows in (**a, c**)). The threshold in bias above which a steep increase in d$I$/d$V$ is observed is approximately 1-2 meV, comparable to the activation energy extracted from the measurements as a function of temperature shown in figure 3.

**Discussion**

It is clear from these observations that the behavior of the gapped insulating state in device A and B is fully analogous, at the quantitative level, with what has been reported earlier for Bernal-stacked 2LG and 4LG [1-5]. In fact, it resembles more the 4LG case, for which the magnitude of the square resistance also coincides quantitatively with that of devices A and B (see figure 3(c)). The observed quantum Hall effect, however, clearly shows that devices A and B are not 4LG (consistently with their larger color contrast in images under an optical microscope). Indeed, for tetralayers the first integer quantum Hall state appearing at low magnetic field is the one at $\nu$ = 8 [1], whereas –as discussed above– devices A and B unambiguously show that the first quantum Hall states that becomes visible are, respectively, at $\nu$ = 12 and 16 (see figure 2). These are the values expected for Bernal-stacked 6LG and 8LG [15]. With this identification, all the measured low-energy electronic properties of device A and B are fully consistent with the initially proposed description of the effect of electron-electron interactions in Bernal-stacked graphene multilayers.

What is truly remarkable is that in all even multilayer devices that we have investigated, with *N*=2, 4, 6, and 8, the insulating state occurs in the same density range and with the same energy gap. This behavior provides a direct experimental indication that the origin of the phenomenon is the same irrespective of the different multilayer thickness, and very strongly suggests that the key aspects of our phenomenological description of interaction effects are correct. It is also remarkable that the magnitude of the square resistance at charge neutrality (at a given temperature) is the same for *N*=4, 6, and 8 (see figure 3(c)), i.e. it is not decreasing for thicker multilayers, from which we establish that the gapped insulating state is not "weakening" as thickness is increasing. We conclude that the simple phenomenological model discussed in the introduction correctly captures the main features of electron-electron interactions in multilayer graphene at least up to 8LG, and that the behavior of tetralayer graphene reported earlier –as well as the inferred even-odd effect– is not coincidental [1].

Both the experimental observations and the validity of the physical scenario accounting for them are worthy of comment. To start with, the staggered layer potential that we introduced is meant to generalize to thicker multilayers the mean-field description that has been developed for bilayers [3, 18, 19, 22]. It should be realized, however, that the validity of this generalization is by far not *a priori* obvious, since in thicker multilayers interactions could have different effects. For instance, in 4LG or thicker even *N* multilayers –but not in bilayers– interactions could mix the different parabolic bands present in a given valley, and prevent the opening of a gap. Additionally, in 4LG or thicker multilayers, effects of interactions associated with layer polarization can play a role (see supporting information). Therefore, even though the phenomenological model that we have proposed, inspired by the theoretical analysis of bilayers, reproduces remarkably well all key experimental observations, a complete understanding will require a detailed microscopic theory to be developed.

The inclusion of more tight-binding couplings, such as those of the Slonczewski-Weiss-McClure model of bulk graphite [10, 34-36] (see figure 1(b)), should also be considered. We find that the majority of these couplings fail to produce new qualitative features relevant to our experiment, different from those already discussed for

bilayers (such as the effect of trigonal warping caused by $\gamma_3$ [20, 23, 37-40]). Specifically, parameter $\delta$, describing a difference in energy between those sites that have neighboring sites directly above or below them and those sites which do not, and parameters $\gamma_3$ and $\gamma_4$, describing skew interlayer hopping, appear solely within the effective Hamiltonians for each set of bilayer-like bands. As a result, they do not mix different sets of bands or prevent the opening of a gap by the staggered potential. Parameters $\gamma_2$ and $\gamma_5$ describe coupling between next-nearest layers and *do* mix different sets of bilayer-like bands. $\gamma_5$, however, affects bands that are at relatively high energy and, thus, it doesn't influence the opening of a gap. Parameter $\gamma_2$, on the contrary, can affect the low-energy part of the band structure, as it tends to mix the bilayer bands and to create a band overlap [10].

Within the scenario that we propose, consistency with the observation of an insulating state in even *N* multilayers, and with the behavior of the low-field integer quantum Hall effect, sets a limit on the magnitude of $\gamma_2$ as being –at most– of the order of the energy gap (namely, smaller than a few meV). The commonly accepted value of $\gamma_2$ from studies of graphite is -20 meV [10, 36] –although estimates range from -20 to +20 meV [10]– but its determination is challenging as well-documented in past literature. In fact, as there are seven parameters in the Slonczewski-Weiss-McClure model, it can be difficult to extract their values independently from a single experiment. Moreover, it has been found that agreement between separate experiments is hampered because of a range of samples and experimental techniques probing different parameter values (energy, magnetic field, etc.) [10]. Even investigations on high quality exfoliated graphene multilayers on-substrate [26], which are certainly more homogeneous than bulk graphite crystals, cannot easily solve the problem, since the interaction with the substrate itself introduces unknown modifications to the potentials of the bottom graphene layers. These potentials enter the equations for the multilayer band structure in a way similar to $\gamma_2$, making it difficult to isolate the effect solely due to this parameter (note, additionally, that devices on a substrate operate in a different regime as compared to our suspended devices because the charge inhomogeneity –larger than $10^{10}$ cm$^{-2}$ even in the best case– is too large for interactions to become dominant). Recently, authors have adopted a variety of procedures when comparing their experimental data with theory. Some have extracted the value of $\gamma_2$ independently [25, 26]; others used the accepted value of $\gamma_2$ [28, 41-43]; others still are able to ignore $\gamma_2$ completely and successfully fit their data to the predictions of the minimal model or a variation thereof [44-47].

At a more fundamental level, however, a key issue is whether the value of the $\gamma_2$ parameter should be the same in even *N* multilayers as it is in odd *N* multilayers or bulk graphite. We think that it may not, due to differences in point group symmetry (between even and odd layers), the possibility of strain or of renormalization effects due to interactions that may be particularly relevant in the regime of the current experiments (in this same regime, the Fermi velocity in monolayers –i.e., the parameter $\gamma_0$– is indeed very strongly renormalized by interactions [8, 48]). To illustrate this point qualitatively, in our theoretical analysis (see supporting information) we consider a specific mechanism and show that in even *N* multilayers, a finite $\gamma_2$ parameter results –at charge neutrality– in a difference in carrier density between layers (i.e., in the presence of finite $\gamma_2$, even *N* multilayers exhibit layer polarization). Maintaining such a difference in charge density requires a large amount of Coulomb energy and is energetically unfavorable, so that the multilayer is likely to "deform" to strongly suppress $\gamma_2$. We

suggest that this occurs through the introduction of interlayer strain which causes the even *N* multilayers to dimerize, i.e., in even-*N* multilayers, the individual layers are slightly displaced, and pair up into *N/2* bilayers with an increased next-nearest layer separation. Such an effect can reduce significantly the magnitude of the $\gamma_2$ parameter, thereby diminishing the energy cost of the layer polarization, leaving all other key parameters of the model ($\gamma_0$ and $\gamma_1$) virtually unchanged. Note that this effect is specific to relatively thin multilayers, since, for graphite the polarization effect is negligible because the density corresponding to each band is spread uniformly over all layers (the polarization effect is thus nearly a surface effect in graphite; see supporting information). This implies that the effect of layer polarization may not only explain why the value of $\gamma_2$ in thin graphene multilayers is much smaller than in graphite, but also why for even *N* multilayers much thicker than 8LG a semi-metallic state analogous to that of graphite is eventually recovered.

**Conclusions**

Besides being striking in its own right, the observation of an insulating state in multilayers as thick as 6LG and 8LG puts in evidence aspects of graphene-based systems that had not been truly appreciated so far. The relevance of layer polarization –which plays no role in mono, bi and trilayers, and had not been considered previously in thicker multilayers– provides one example. The large degeneracy of electronic states at the K and K' points provides another one. It has been long known that within the minimal tight binding model with only $\gamma_0$ and $\gamma_1 \neq 0$ all quadratic bands in even *N* multilayers are degenerate at the K (K') point [11-17]. It was believed, however, that such a degeneracy would be lifted in practice. The observed experimental behavior, however, shows that this large degeneracy is not only actually present, but that it even grows with increasing layer thickness: the occurrence of the lowest energy integer quantum Hall effect state at *ν* = 4, 8, 12, and 16 for 2LG, 4LG, 6LG, and 8LG is a direct manifestation of this fact. Such a degeneracy can lead to new physical phenomena. That is the case, for instance, when entering the fractional quantum Hall regime, since in thick even *N* multilayers interactions can mix a large number of *E* = 0 degenerate Landau levels, distinct only for their different orbital quantum number. Such a regime is not accessible in conventional GaAs-based two-dimensional electron gases [49-56]. It is in graphene bilayers, where two such orbital *E* = 0 levels are present and new unusual behavior and even denominator fractional states have indeed been observed [32, 57-59]. It should be expected to have even more drastic consequences in thicker even *N* multilayers, as the degeneracy of the *E* = 0 state is larger.

It seems hard to imagine that such a large degeneracy is coincidental and –since the degeneracy does not appear to be protected by any known symmetry of the material structure or by time reversal symmetry– one is left wondering about the physical phenomenon behind its occurrence. More in general, finding robust interaction-driven phenomena that exhibit a behavior as systematic as the one that we have reported here is rare. Both the robustness and the systematics call for a detailed microscopic theoretical analysis, which is essential to justify the phenomenological model that we have proposed, and to understand why it works as well as it does. We anticipate that such an analysis may reveal more aspects of the electronic properties of graphene-

based systems that had not been appreciated until now. It is certainly remarkable that, despite the decade of very intense research on graphene-based systems and many decades of work on graphite, these systems continue to reveal unexpected and surprising phenomena.


**Acknowledgement**

We gratefully acknowledge A. Ferreira for technical help. The Swiss National Science Foundation, the Center of Excellence in Research NCCR QSIT, and the EU project "Graphene Flagship" are also very gratefully acknowledged for financial support.

# Supporting information

The material presented in this Supporting information serves a twofold purpose. For the convenience of the reader, the first goal is to present in more detail the theoretical scenario that we have used to explain the experimental data, *i.e.* the minimal tight-binding model augmented by a staggered potential. The second –and more interesting– goal is to introduce the concept of layer polarization, which has not been considered previously in the context of graphene multilayers, and to discuss qualitatively its relevance. As we will argue, avoiding the electrostatic energy cost associated to layer polarization can explain why the parameter $\gamma_2$ should be expected to be strongly suppressed in Bernal multilayers as compared to graphite. The related physics seems also to be relevant to determine the thickness at which multilayers should start to behave as bulk graphite, and to understand the disappearance of the insulating state at finite carrier density. We make clear from the start that our discussion is not meant to be a formal and complete theoretical treatment of the problem. It only aims at pointing to a robust –and seemingly very relevant– physical phenomenon, which appears to support the phenomenological model that we have used in the main text. All our considerations are summarized in the Conclusions Section at the end.

## I. PHENOMENOLOGICAL MODEL OF MULTILAYER GRAPHENE

In the following we describe our simple phenomenological approach, which is based on the so-called minimal tight-binding model of multilayer graphene augmented with a staggered layer potential [1]. The minimal tight-binding description includes only nearest-neighbor intra- and inter-layer coupling (usually denoted by parameters $\gamma_0$ and $\gamma_1$), the staggered layer potential is $V_i = (-1)^{(i+1)}\Delta$ where $i = 1, 2, 3, \ldots, N$ is a layer index and $\Delta$ is the order parameter, assumed to originate from a mean-field treatment of electron-electron interactions.

We consider Bernal-stacked multilayer graphene [2, 3] with spatial Cartesian components $(x, y, z)$, where $x, y$, are in the graphene plane, $z$ is in the direction of stacking. Primitive lattice vectors are $\mathbf{a}_1$, $\mathbf{a}_2$ and $\mathbf{a}_3$:

$$\mathbf{a}_1 = \left(\frac{a}{2}, \frac{\sqrt{3}a}{2}, 0\right), \qquad \mathbf{a}_2 = \left(\frac{a}{2}, -\frac{\sqrt{3}a}{2}, 0\right), \qquad \mathbf{a}_3 = (0, 0, 2c),$$

where $a = |\mathbf{a}_1| = |\mathbf{a}_2|$ is the lattice constant in the plane, $c$ is the interlayer distance, and corresponding primitive reciprocal lattice vectors $\mathbf{b}_1$, $\mathbf{b}_2$ and $\mathbf{b}_3$ are

$$\mathbf{b}_1 = \left(\frac{2\pi}{a}, \frac{2\pi}{\sqrt{3}a}, 0\right), \qquad \mathbf{b}_2 = \left(\frac{2\pi}{a}, -\frac{2\pi}{\sqrt{3}a}, 0\right), \qquad \mathbf{b}_3 = \left(0, 0, \frac{\pi}{c}\right).$$

Bloch functions are

$$\Phi_j(\mathbf{k}, \mathbf{r}) = \frac{1}{\sqrt{M}} \sum_{i=1}^{M} e^{i\mathbf{k}\cdot\mathbf{R}_{j,i}} \phi_j(\mathbf{r} - \mathbf{R}_{j,i}),$$

where the sum is over $M$ different unit cells, labeled by index $i = 1 \ldots M$, and $\mathbf{R}_{j,i}$ denotes the position of the $j$th orbital in the $i$th unit cell. In each unit cell there are four atoms, $A_n$, $B_n$, $A_{n+1}$, $B_{n+1}$, for the $n$th and $(n+1)$th layers, we take into account one $p_z$ orbital per site. In the tight-binding model, we determine transfer integral matrix elements $H_{il} = \langle \Phi_i | \mathcal{H} | \Phi_l \rangle$ to give the following Schrödinger equations for each atom:

$$v\pi^\dagger B_n = (E - \Delta)A_n,$$
$$v\pi A_n + \gamma_1 A_{n+1} + \gamma_1 A_{n-1} = (E - \Delta)B_n,$$
$$v\pi^\dagger B_{n+1} + \gamma_1 B_{n+2} + \gamma_1 B_n = (E + \Delta)A_{n+1},$$
$$v\pi A_{n+1} = (E + \Delta)B_{n+1},$$

where $\pi = \xi p_x + i p_y$, $\pi^\dagger = \xi p_x - i p_y$, $\xi = \pm 1$ is a valley index, and $v = \sqrt{3}a\gamma_0/(2\hbar)$. These equations are solved using plane waves with wave vector $k_z$ in the stacking direction,

$$A_n = A_1 e^{ik_z nc},$$
$$B_n = B_1 e^{ik_z nc},$$
$$A_{n+1} = A_2 e^{ik_z(n+1)c},$$
$$B_{n+1} = B_2 e^{ik_z(n+1)c},$$



so that, in basis $A1, B1, A2, B2$, the Hamiltonian is

$$H = \begin{pmatrix} \Delta & v\pi^\dagger & 0 & 0 \\ v\pi & \Delta & \gamma & 0 \\ 0 & \gamma & -\Delta & v\pi^\dagger \\ 0 & 0 & v\pi & -\Delta \end{pmatrix}, \quad (1)$$

$$\gamma = 2\gamma_1 \cos(k_z c).$$

Eigenvalues [4] are given by

$$E_{s,u} = s\left[\frac{\gamma^2}{2} + \Delta^2 + \hbar^2 v^2(k_x^2 + k_y^2) + u\sqrt{\frac{\gamma^4}{4} + \hbar^2 v^2(k_x^2 + k_y^2)(\gamma^2 + 4\Delta^2)}\right]^{1/2}, \quad (2)$$

where $s = \pm 1$ denotes conduction/valence bands and $u = \pm 1$ denotes high- and low-energy states.

For a finite number of layers $N$, we consider the vertical $z$ component to run from $z = c$ (the bottom layer) to $z = Nc$ (the top layer). The upper two components $A1, B1$ of the Hamiltonian (1) correspond to odd layers $z = c, 3c, 5c, \ldots, (N-1)c$ and the lower two $A2, B2$ correspond to even layers $z = 2c, 4c, 6c, \ldots, Nc$. Boundary conditions state that the wave functions will be zero outside the system, at fictitious layers $z = 0$ and $z = c(N+1)$. Using these conditions gives a standing (sine) wave and it quantizes the vertical wave vector [5] as

$$k_z = \frac{\ell \pi}{c(N+1)}, \quad (3)$$

where the mode index $\ell = 1, 2, 3, \ldots N/2$ for even $N$ and $\ell = 1, 2, 3, \ldots (N+1)/2$ for odd $N$. This quantized $k_z$ values enter the dispersion relation (2) through the parameter $\gamma$, and determine the properties of the electronic bands. Most of these bands have a quadratic dispersion relation as in graphene bilayer, except when $N$ is odd and $\ell = (N+1)/2$, which yields $k_z c = \pi/2$ and $\gamma = 0$, corresponding to linearly dispersing monolayer-like bands.

Interestingly, the monolayer-like bands remain gapless in the presence of finite $\Delta$ because they have finite amplitude on every other layer (the odd layers at $z = c, 3c, 5c, \ldots, (N-1)c$) and, thus, experience a constant layer potential. This is in a clear contrast with the behavior of bilayer-like bands that are gapped by the staggered layer potential $\Delta$ as sketched in Fig. 1c in the main text, leading to the observed insulating behavior of even-$N$ multilayers. Note also that in the quantum-Hall regime, each monolayer and bilayer-like bands give rise to 4-fold and 8-fold degenerate zero energy Landau level, respectively. The model therefore predicts the first integer quantum-Hall plateau to occur at $\nu = 2 \times N$ in $N$ multilayer graphene as observed experimentally. In summary, this phenomenological "minimal" model including only the tight-binding parameters $\gamma_0$ and $\gamma_1$, augmented by a staggered potential, does correctly capture the insulating behavior of even-$N$ multilayer graphene, the fact that odd-$N$ multilayers stay conducting at low temperature with a conductivity of a few $e^2/h$, as well as of the filling factor at which the first integer quantum-Hall effect appears.

## II. LAYER POLARIZATION

In the main text, we mentioned that there exists a finite effect of layer polarization in thin graphene multilayers which can lead to the interaction-driven suppression of $\gamma_2$ at charge neutrality, as it is needed to justify the application of the minimal tight-binding model in our phenomenological models discussed above. Here we discuss the origin of layer polarization in more detail.

To describe the layer polarization effect in multilayer graphene, we consider wave functions corresponding to Hamiltonian (1) with quantized wave vector (3). For simplicity, we set $\Delta = 0$ (this will not affect the main conclusion in a qualitative way). Then, in general, the wave function may be written as

$$\psi_{\ell,u,s}(\mathbf{r}) = \frac{1}{\sqrt{(E_{s,u}^2 + \hbar^2 v^2[k_x^2 + k_y^2])}} \begin{pmatrix} \hbar v(\xi k_x - i k_y) \\ E_{s,u} \\ -us E_{s,u} \\ -us \hbar v(\xi k_x + i k_y) \end{pmatrix} \sqrt{\frac{2}{c(N+1)}} \sin\left[\frac{\pi \ell z}{(N+1)c}\right] \frac{e^{i(k_x x + k_y y)}}{\sqrt{A}},$$

where $A$ is the area per graphene plane, used to normalize plane waves with periodic boundary conditions in the $x$-$y$ plane. Note that for each $\ell$ value, there are four solutions $E_{s,u}$ due to $u = \pm 1$, $s = \pm 1$. An exception to this is the



monolayer-like mode for odd $N$, $k_z c = \pi/2$ for which there are two solutions, $E_s = svp$, with eigenstates:

$$\psi_s(\mathbf{r}) = \frac{1}{\sqrt{(E_s^2 + \hbar^2 v^2[k_x^2 + k_y^2])}} \begin{pmatrix} \hbar v(\xi k_x - i k_y) \\ E_s \\ 0 \\ 0 \end{pmatrix} \sqrt{\frac{2}{c(N+1)}} \sin\left[\frac{\pi z}{2c}\right] \frac{e^{i(k_x x + k_y y)}}{\sqrt{A}},$$

indicating that the wave function only has finite amplitude on every other layer [the odd layers at $z = c, 3c, 5c, \ldots, (N-1)c$].

Using the wave functions, it is possible to determine the probability density for each layer of the multilayer. For each state, the probability density per layer is given by:

$$\phi_\ell(z) = \frac{2}{cA(N+1)} \sin^2\left[\frac{\pi \ell z}{(N+1)c}\right].$$

For even $N$, if we sum over all states in the valence band, this requires summation over all $\ell$ and a factor of two for the index $u = \pm 1$, giving the total probability density per layer:

$$\phi(z) = \frac{4}{cA(N+1)} \sum_{\ell=1}^{N/2} \sin^2\left[\frac{\pi \ell z}{(N+1)c}\right]. \tag{4}$$

For odd $N$, if we sum over all states in the valence band, this requires summation over all $\ell$ and a factor of two for the index $u = \pm 1$ for all states except the monolayer one, giving the total probability density per layer:

$$\phi(z) = \frac{4}{cA(N+1)} \left\{ \sum_{\ell=1}^{(N-1)/2} \sin^2\left[\frac{\pi \ell z}{(N+1)c}\right] + \frac{1}{2} \sin^2\left[\frac{\pi z}{2c}\right] \right\}.$$

In both cases, even and odd $N$, the probability density on each layer (i.e. where $z$ is an integer multiple of $c$) simplifies as $\phi(z) = 1/(cA)$. The probability density takes the same value on every layer, which is to be expected as there is a summation over every state in the valence band. To determine the carrier density $n$ per layer, one should integrate with respect to the wave vector in the graphene plane $\mathbf{k} = (k_x, k_y)$. For example, for even $N$,

$$n = \frac{8}{\pi(N+1)} \sum_{\ell=1}^{N/2} \int_{\text{occ}} k dk \sin^2\left[\frac{\pi \ell z}{(N+1)c}\right], \tag{5}$$

where we introduced a factor of 4 for valley and spin degeneracy, and the integral is taken over all occupied $k = |\mathbf{k}|$ values. When the Fermi level is at zero energy, the valence band is fully occupied and the integral $\int_{\text{occ}} k dk$ takes the same value for all states: just as with the probability density, the density on each layer is the same. However, for finite doping, the occupied states of different bands have different ranges of $k$ values and, as a consequence, the integrals in Eq. (5) take different values for each $\ell$: when summing over all the bands (the $\ell$ values), the density $n$ now acquires a layer dependence.

In order to illustrate the layer dependence of the polarization effect in the simplest possible way, instead of attempting to evaluate integrals $\int_{\text{occ}} k dk$ for individual bands, we take into account only a certain fraction of the states in the summation in Eq. (4) to consider the probability density per layer. For example, Fig. 1 shows the probability density taking into account only half of the states in the valence band, $\ell = 1, 2, \ldots N/4$, in Eq. (4). The plots for thin multilayers, $N = 4$ and $N = 8$, show a large difference in the density between adjacent layers, this is the layer polarization effect, although spatial inversion (reflection with respect to $z$ about the mid-point in the figures) of the multilayer is preserved. The plot for a thick multilayer ($N = 800$) shows only one end of the multilayer (the first 100 layers) and it illustrates that the layer polarization can be considered as a surface effect of bulk graphite. This plot suggest that the layer polarization effect decays on a length scale of roughly 20 layers. In this section, we considered layer polarization created by an incomplete filling of the bands, in the next section we consider the role of next-nearest layer coupling.

### III. THE ROLE OF NEXT-NEAREST LAYER COUPLING

In the previous section, we described how layer polarization may result from incomplete filling of the bands in multilayer graphene. Here, we illustrate that the polarization effect may also occur due to next-nearest layer coupling,

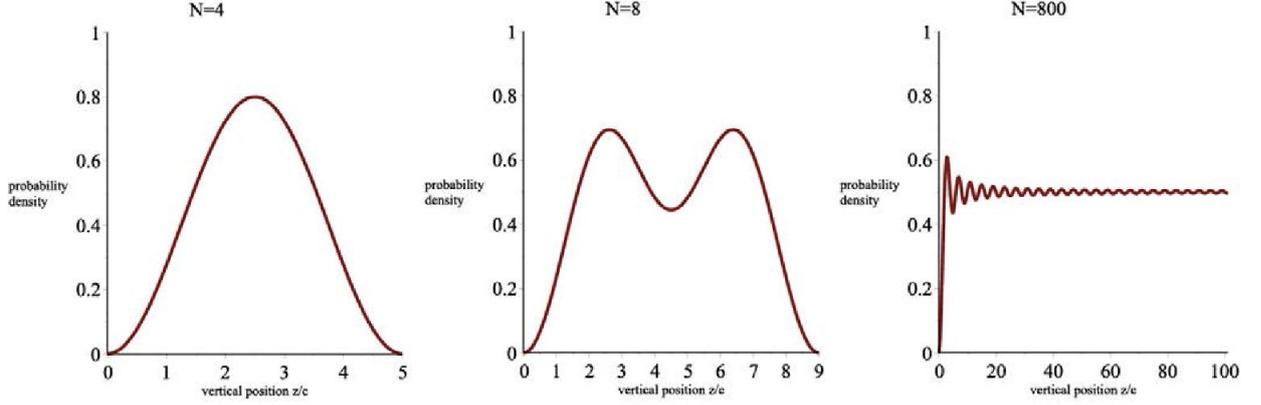

FIG. 1: The probability density per layer for multilayer graphene with $N$ layers, taking into account only half of the states, $\ell = 1, 2, \ldots N/4$, in the valence band, see Eq. (4). Note that sites $z = 0$ and $z = (N+1)c$ are outside the multilayer, where the wave function is defined to be zero, $c$ is the interlayer separation. For the thick multilayer, $N = 800$, only one end of the multilayer is plotted (the first 100 layers).

as parameterized by $\gamma_2$. To describe this effect, we concentrate, for simplicity, on hexalayers (although the same arguments will apply to thin multilayers including tetra- and octalayer graphene).

The unit cell of hexalayer graphene is shown schematically on the left side of Fig. 2(a), and we take into account one $p_z$ orbital per site. There are twelve atomic sites in the unit cell, two per layer. Half of the sites ($B1$, $A2$, $B3$, $A4$, $B5$, $A6$) are connected to an atom above and/or below by relatively strong interlayer coupling $\gamma_1$ (indicated by a solid vertical line in Fig. 2(a)). These sites are called 'dimer' sites because, owing to the interlayer coupling, the states corresponding to these atoms are split away from zero energy by an energy of the order of $\gamma_1$. Thus, for energies smaller than $\gamma_1$, an effective Hamiltonian may be derived by eliminating the components related to the dimer sites (generalizing a procedure developed for bilayer graphene [4]). Then, it is possible to consider a Hamiltonian $H_6$ written explicitly in a basis consisting of a single orbital on each layer, on non-dimer sites $A1$, $B2$, $A3$, $B4$, $A5$, $B6$, which is valid at low energy only:

$$H_6 = \begin{pmatrix} \Delta & -(v\pi^\dagger)^2/\gamma_1 & \gamma_2/2 & (v\pi^\dagger)^2/\gamma_1 & 0 & -(v\pi^\dagger)^2/\gamma_1 \\ -(v\pi)^2/\gamma_1 & -\Delta & 0 & \gamma_2/2 & 0 & 0 \\ \gamma_2/2 & 0 & \Delta & -(v\pi^\dagger)^2/\gamma_1 & \gamma_2/2 & (v\pi^\dagger)^2/\gamma_1 \\ (v\pi)^2/\gamma_1 & \gamma_2/2 & -(v\pi)^2/\gamma_1 & -\Delta & 0 & \gamma_2/2 \\ 0 & 0 & \gamma_2/2 & 0 & \Delta & -(v\pi^\dagger)^2/\gamma_1 \\ -(v\pi)^2/\gamma_1 & 0 & (v\pi)^2/\gamma_1 & \gamma_2/2 & -(v\pi)^2/\gamma_1 & -\Delta \end{pmatrix}. \quad (6)$$

The top left $2 \times 2$ block of Eq. (6) describes a bilayer consisting of layers 1 and 2, the central $2 \times 2$ block describes a bilayer consisting of layers 3 and 4, and the bottom right $2 \times 2$ block describes a bilayer consisting of layers 5 and 6. Off-diagonal blocks account for coupling between them with $\gamma_2$ characterizing next-nearest-layer hopping between layers 1 and 3, layers 2 and 4, layers 3 and 5, and layers 4 and 6. Terms such as $-(v\pi)^2/\gamma_1$ on the off-diagonal describe effective coupling between distance atoms through hopping via dimer sites.

For $\gamma_2 = \Delta = 0$, there are six parabolic bands that touch at zero energy which we label with index $i = 1, 2, 3, 4, 5, 6$ in order of descending energy $E_i = \alpha_i v^2 p^2 / \gamma_1$ where $\alpha_i$ are real numbers: $\alpha_1 \approx 2.247$, $\alpha_2 \approx 0.802$, $\alpha_3 \approx 0.555$, $\alpha_4 = -\alpha_3$, $\alpha_5 = -\alpha_2$, $\alpha_6 = -\alpha_1$. The Hamiltonian satisfies spatial inversion symmetry such that electrons in bands 1, 3 and 5 have odd parity, bands 2, 4 and 6 are even. In basis $A1$, $B2$, $A3$, $B4$, $A5$, $B6$, spinor eigenstates for $\gamma_2 = \Delta = 0$ may be written as

$$\psi_i^{(0)}(\mathbf{r}) = N_i \begin{pmatrix} \alpha_i^4 \\ -\alpha_i^3 e^{2i\xi\phi} \\ \alpha_i^2(1-\alpha_i^2) \\ -\alpha_i(1-2\alpha_i^2)e^{2i\xi\phi} \\ (1-3\alpha_i^2+\alpha_i^4) \\ -\alpha_i(1-3\alpha_i^2+\alpha_i^4)e^{2i\xi\phi} \end{pmatrix} e^{i\mathbf{p}\cdot\mathbf{r}/\hbar}, \quad (7)$$

where $\phi$ is the polar angle of momentum $\mathbf{p}$ and the normalization constants are $N_1 = N_6 \approx 0.0204$, $N_2 = N_5 \approx 1.0105$, $N_3 = N_4 \approx 2.4451$. The eigenstates yield the probability density for each layer $|A1|^2 = |B6|^2$, $|B2|^2 = |A5|^2$,



$|A3|^2 = |B4|^2$ (layers 1 and 6 *etc.* have the same density due to spatial inversion symmetry) meaning that the density is split between the first, second and third layers in the ratio 0.54 : 0.11 : 0.35 for bands 1 and 6, the ratio 0.35 : 0.54 : 0.11 for bands 2 and 5, and 0.11 : 0.35 : 0.54 for bands 3 and 4. In general, this creates a layer polarization effect, although, when the Fermi level is at zero energy there are three completely filled valence bands 4, 5 and 6 for which the layer polarization effect cancels overall with density distributed homogeneously over the layers, $|A1|^2 = |B2|^2 = |A3|^2 = |B4|^2 = |A5|^2 = |B6|^2 = 1/2$ (these sum to three because there are three valence bands).

When the Fermi level is not at zero energy, cancellation of the layer polarization effect by partially filled bands (conduction or valence) is imperfect, as described in the previous section. Even when the Fermi level is at zero, however, the effect of next-nearest layer coupling $\gamma_2$ is to modify the eigenstates and to introduce a finite layer polarization. To demonstrate this, we treat the part of the Hamiltonian containing $\gamma_2$ as a perturbation $\delta H$ and use first-order perturbation theory $\psi_i = \psi_i^{(0)} + \sum_{n \neq i} \langle \psi_n^{(0)} | \delta H | \psi_i^{(0)} \rangle \psi_n^{(0)} / (E_i - E_n)$, valid for $v^2 p^2 \gg \gamma_1 \gamma_2$. As the perturbation $\gamma_2$ arises from coupling between atomic orbitals, it doesn't break the spatial inversion symmetry of the lattice and, thus, can only couple bands of the same parity, namely 1, 3 and 5, and 2, 4 and 6. We find $\psi_4 \approx \psi_4^{(0)} - \lambda \psi_2^{(0)} + \lambda \psi_6^{(0)}$, $\psi_5 \approx \psi_5^{(0)} - \lambda \psi_1^{(0)} - \lambda \psi_3^{(0)}$, $\psi_6 \approx \psi_6^{(0)} - \lambda \psi_2^{(0)} - \lambda \psi_4^{(0)}$ for the valence bands where $\lambda = \gamma_1 \gamma_2 / (14 v^2 p^2)$ (similar expressions hold for the conduction bands). The correction to the layer probability density does not cancel when combining the contribution of valence bands 4, 5 and 6, but it accumulates to give $|A1|^2 = |B6|^2 = 1/2 - 1.2588\lambda$, $|B2|^2 = |A5|^2 = 1/2 + 1.3546\lambda$, and $|A3|^2 = |B4|^2 = 1/2 - 0.0958\lambda$, *i.e.* there is a difference in probability density between the layers. Maintaining a difference in charge density between different layers is energetically unfavorable because it requires a large amount of Coulomb energy, and we believe this is why the value of $\gamma_2$ is suppressed in thin, even-$N$ multilayers. We do not know the microscopic mechanism for such a suppression: it may be that electron interactions lead to a renormalization of the value of $\gamma_2$ (it is known that the Fermi velocity of monolayer graphene is renormalized by interactions [6–10]). As an alternative example, in the next section we describe a simple, qualitative model of interlayer strain and analyze whether this is a suitable candidate.

## IV. QUALITATIVE MODEL OF INTERLAYER STRAIN

Maintaining a difference in charge density between layers is energetically unfavorable because it requires a large amount of Coulomb energy. In this section, we describe a simple model of interlayer strain and discuss whether it is a possible microscopic mechanism for the suppression of the magnitude of next-nearest layer coupling $\gamma_2$. Our model is based on dimerization of the layers: interlayer strain causes layers to pair up into bilayers, *i.e.* layers 1 and 2, 3 and 4, and 5 and 6 move slightly towards each other, forming three bilayers as shown schematically on the right side of Fig. 2(b) with an increased separation (and, hence, increased next-nearest layer separation) between the bilayers. In this case, the value of $\gamma_2$ would be reduced, and the value of $\gamma_1$ would take two possible values: $\gamma_+$ for strong coupling within each bilayer, $\gamma_-$ for relatively weak coupling between bilayers. As a result, the effective Hamiltonian may be written as

$$H_6 = \begin{pmatrix} \Delta & -(v\pi^\dagger)^2/\gamma_+ & \gamma_2/2 & (v\pi^\dagger)^2\gamma_-/\gamma_+^2 & 0 & -(v\pi^\dagger)^2\gamma_-^2/\gamma_+^3 \\ -(v\pi)^2/\gamma_+ & -\Delta & 0 & \gamma_2/2 & 0 & 0 \\ \gamma_2/2 & 0 & \Delta & -(v\pi^\dagger)^2/\gamma_+ & \gamma_2/2 & (v\pi^\dagger)^2\gamma_-/\gamma_+^2 \\ (v\pi)^2\gamma_-/\gamma_+^2 & \gamma_2/2 & -(v\pi)^2/\gamma_+ & -\Delta & 0 & \gamma_2/2 \\ 0 & 0 & \gamma_2/2 & 0 & \Delta & -(v\pi^\dagger)^2/\gamma_+ \\ -(v\pi)^2\gamma_-^2/\gamma_+^3 & 0 & (v\pi)^2\gamma_-/\gamma_+^2 & \gamma_2/2 & -(v\pi)^2/\gamma_+ & -\Delta \end{pmatrix}. \quad (8)$$

Effective couplings between distant atoms that are quadratic in momentum [*e.g.* the terms $(v\pi)^2\gamma_-/\gamma_+^2$ and $-(v\pi)^2\gamma_-^2/\gamma_+^3$] are smaller than the coupling within each bilayer [*e.g.* the term $-(v\pi)^2/\gamma_+$] because of the small parameter $\gamma_-/\gamma_+ < 1$. As an illustrative example, Fig. 2(b)-(d) shows the low-energy bands in the presence of non-zero $\Delta$. Panel (b) shows the bands for $\gamma_2 = 0$, in this case a clear gap is seen and the bands are degenerate at small momentum $p$. The effect of finite $\gamma_2$, however, is to close the gap and to create splitting between the bands at small $p$, as shown in panel (c). Reducing the value of $\gamma_2$ through the presence of strain causes the gap to appear again, panel (d), albeit with a smaller value than in (b), and the splitting between the bands at small $p$ is reduced, too. The calculation described [as shown in Fig. 2(d)] would not give the correct sequencing of plateaus in the quantum Hall effect because the degeneracy of the parabolic bands at small $p$ is still broken. However, Fig. 2(d) is just an illustration to demonstrate how the value of $\gamma_2$ could be suppressed. If the suppression was large enough, the degeneracy of the bands at small $p$ would be recovered [as in Fig. 2(b)]. In other words, in the limit that strain suppresses the value of $\gamma_2$ completely, the net effect is a gapped spectrum with degenerate bands at low energy but with different values of effective mass [as compared to those in Fig. 2(b)] due to the dimerization of the values of $\gamma_1$ [as described by Eq. (8)].

Fig. 2 illustrates an example of how the influence of next-layer coupling may be mitigated by interlayer strain, but we would like to stress that we do not have a quantitative, microscopic theory of this effect. For example, in

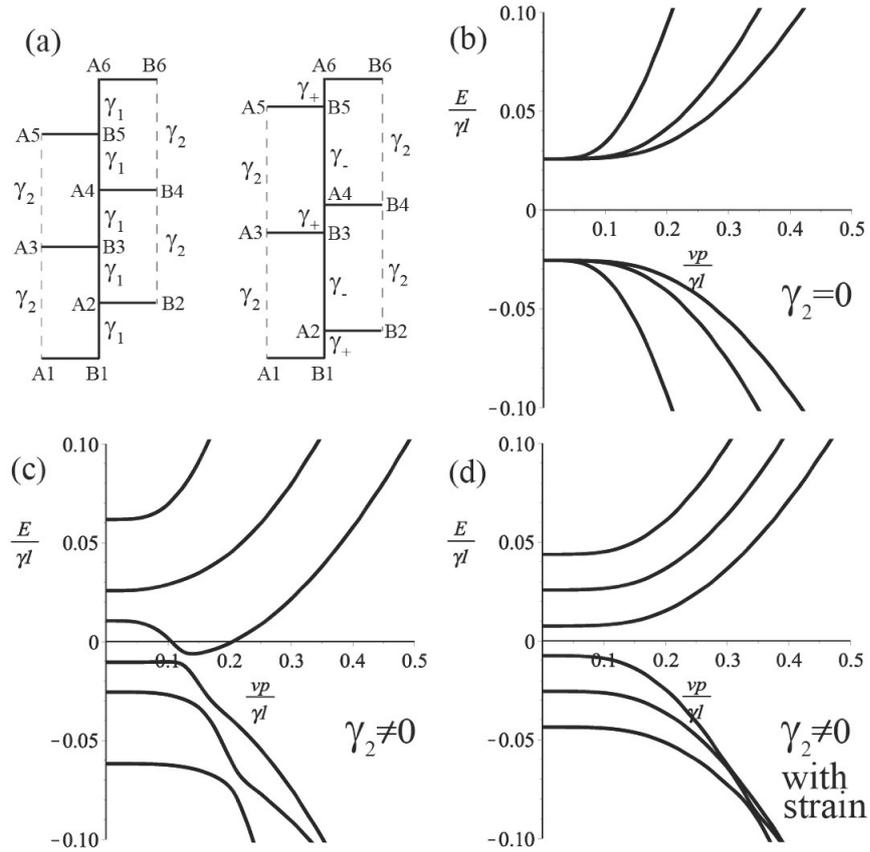

FIG. 2: (a) Schematic of the unit cell of hexalayer graphene (left) with twelve atomic sites, two per layer. Solid vertical lines show interlayer coupling $\gamma_1$, dashed vertical lines show next-nearest layer coupling $\gamma_2$. The right side shows an exaggerated view of how interlayer strain could reduce the magnitude of next-nearest layer coupling $\gamma_2$, by pairing the layers into bilayers. The value of $\gamma_1$ can then take two possible values: $\gamma_+$ for strong coupling within each bilayer, $\gamma_-$ for relatively weak coupling between bilayers. (b), (c), (d) show the low-energy bands of hexalayer graphene for $\gamma_1 = 0.39\,\text{eV}$ and $\Delta = 10\,\text{meV}$. Panel (b) shows the bands for $\gamma_2 = 0$, (c) for $\gamma_2 = -20\,\text{meV}$, obtained from Hamiltonian (6). Panel (d) shows the effect of strain for finite $\gamma_2$, obtained from Hamiltonian (8), with parameters $\gamma_-$ and $\gamma_2$ reduced by 50%, parameter $\gamma_+$ increased by 50%.

Fig. 2(d) we have altered all interlayer parameters by the same amount, 50%, for illustrative purposes, but we believe that parameter $\gamma_2$ will be particularly sensitive to a modification of the interlayer spacing because it describes *next-nearest-layer* coupling (as opposed to $\gamma_1$ which describes nearest-layer coupling). If the characteristic length scale for next-nearest layer coupling is of the order of the Bohr radius $a_B \approx 0.529\,\text{Å}$, then a next-nearest-layer coupling parameter at a distance of two interlayer spacings$\approx 6.70\,\text{Å}$ is already in the rapidly-decaying tail of the spatial dependence and any further increase in separation will quickly suppress its magnitude. For example, an exponential tail $\exp(-r/a_B)$ of the spatial dependence would mean that only a 5% increase in the interlayer separation produces a 50% fall in the value of $\gamma_2$.

In summary, our low-energy minimal tight-binding model of hexalayer graphene with a staggered potential produces a gap and degenerate bands at small momentum $p$, Fig. 2(b). Introducing next-nearest layer coupling $\gamma_2$ has the effect of closing the gap and breaking the degeneracy of the bands, but a particular model of deformation of the lattice, namely, dimerization of the layers caused by interlayer strain, is able to strongly suppress the value of $\gamma_2$, thus opening the gap and restoring the degeneracy of the bands [Fig. 2(d) shows an example, although here the suppression of $\gamma_2$ is not complete]. Note that dimerization of the nearest-neighbor interlayer parameter $\gamma_1$ leads to a change in the effective masses of the bands at low energy, but it does not break their degeneracy (only $\gamma_2$ splits the bands).

## V. CONCLUSIONS

While clearly not representing a comprehensive and rigorous analysis, the theoretical considerations presented above illustrate aspects of the physics that are important to understand our experimental results on thick Bernal multilayers, as well as the phenomenological model that we have introduced to explain them. Specifically, they reveal the role of layer polarization as a phenomenon that –despite having clear relevance to understand the electronic properties of these systems– had not been previously considered in discussions. Here we summarize the key points that result from our theoretical considerations, and emphasize explicitly how they support the physical scenario that we have used in the main text to interpret our experiments.

A key observation is that, at charge neutrality (or very close to the charge neutrality point), a finite value for the parameter $\gamma_2$ causes the occurrence of layer polarization, *i.e.* the presence of a non-uniform, layer-dependent electron density in even multilayers starting from 4LG. Layer polarization costs electrostatic energy, and is energetically unfavorable. It should therefore be expected that, to minimize its electrostatic energy cost, the system will "deform" in such a way to strongly suppress $\gamma_2$. In the sections above, we have discussed a specific type of deformation –namely a dimerization of the layers– that can lead to a large suppression of $\gamma_2$ without changing the physics captured by the phenomenological model introduced in the main text. Even though other examples of deformation can likely be found, and a quantitative analysis will then be needed to understand which one occurs in practice, the main point to be retained is that a realistic physical scenario exists which leads to the suppression of $\gamma_2$ without altering the physics of interest.

Irrespective of the specific "deformation" process, we then conclude that $\gamma_2$ should be expected to be much smaller in thin multilayers as compared to bulk graphite. Indeed, our theoretical discussion shows that inside the bulk of a thick graphite crystal layer polarization vanishes irrespective of the value of $\gamma_2$. In other words, inside the bulk of graphite the driving force to suppress $\gamma_2$ is not present, so that no dimerization (or any other type of "deformation") is expected. Clearly this implies that the phenomenon of layer polarization can potentially account for why bulk graphite and thinner multilayers behave so differently.

On this same topic, our analysis shows that layer polarization –while being absent in the bulk– does persist near the surface of bulk graphite. From the simple calculations presented here, it seems that the depth of the surface region with non-vanishing layer polarization is rather extended, as much as several tens of layers. Clearly, a precise and reliable determination of the depth over which layer polarization is present requires a quantitative, detailed theoretical study. Nevertheless, our results strongly suggest that the depth of the surface region is considerably longer than the thickness of all suspended multilayers that have been investigated so far. This is important, as it indicates that layer polarization may be the phenomenon determining the length scale at which the crossover from thick multilayer to bulk graphite occurs. In other words, finding that layer polarization decays away from the surface only very slowly is compatible with our experimental observation that the multilayers that we have investigated exhibit a behavior drastically different from that of bulk graphite, albeit they would be normally considered to be rather thick.

Finally, we argue that layer polarization may also play a role in determining the narrow density range in which the insulating state in suspended multilayers is observed. That is because away from charge neutrality layer polarization is present irrespective of whether $\gamma_2$ vanishes or not, so that at sufficiently large density there is no drive to "deform" the system to suppress $\gamma_2$. What this means physically is that $\gamma_2$ should be expected to depend on the density of charge carriers $n$, *i.e.*, as a result of electron-electron interactions $\gamma_2$ is renormalized in a way that parallels what happens to $\gamma_0$ [6–10]. Since a larger, unsuppressed $\gamma_2$ parameter results in an overlap between valence and conduction bands, such a scenario would explain in an internally consistent way why, at finite carrier density, all even multilayers become again metallic and the gap disappears.

We emphasize once again that the considerations presented here are qualitative and, by themselves, cannot determine the order of magnitude of the effect discussed (*i.e.*, the dependence of $\gamma_2$ on $n$). They nevertheless point to a clear scenario rooted in a robust physical phenomenon (avoiding deviations from local charge neutrality) which supports the validity of the phenomenological theoretical model discussed in the main text (the minimal tight binding model complemented by a staggered potential, which accounts for our experimental observations). Both the quite remarkable behavior of thick Bernal multilayers and the theoretical considerations presented here provide a clear motivation to carry out a systematic theoretical analysis of electron-electron interactions in these systems, and indicate that –in such an analysis– the concept of layer polarization is likely to play a prime role.

---